\documentclass[acmtog,dvipsnames,screen,nonacm]{acmart}  %

\usepackage{tabularx,booktabs} %
\usepackage[normalem]{ulem}  %
\usepackage{color,hyphenat,balance}
\usepackage[fleqn,tbtags]{mathtools}
\usepackage{overpic}  %
\usepackage{wrapfig}
\usepackage{contour}
\usepackage{microtype}
\usepackage{multirow}
\usepackage{amsmath}
\usepackage{bbm}
\usepackage{colortbl}
\usepackage{textcomp}
\usepackage{enumitem}
\setitemize{noitemsep,topsep=0pt,parsep=0pt,partopsep=0pt}
\usepackage{derivative}
\usepackage{float}
\usepackage[most]{tcolorbox}
\usepackage{graphicx}
\graphicspath{{./images/}}
\usepackage{listings}
\usepackage{amsmath,amsfonts,bm}

\def\1{\bm{1}}

\DeclareMathAlphabet{\mathsfit}{\encodingdefault}{\sfdefault}{m}{sl}
\SetMathAlphabet{\mathsfit}{bold}{\encodingdefault}{\sfdefault}{bx}{n}

\def \etal {{\emph{et al}.\thinspace}}
\def \eg {{\emph{e.g.},\thinspace}}
\def \ie {{\emph{i.e.},\thinspace}}

\usepackage[ruled]{algorithm2e} 

\SetAlFnt{\small}
\SetAlCapFnt{\small}
\SetAlCapNameFnt{\small}
\SetAlCapHSkip{0pt}

\def \etal {{\emph{et al}.\thinspace}}
\def \eg {{\emph{e.g.},\thinspace}}
\def \ie {{\emph{i.e.},\thinspace}}

\newcommand{\rev}[1]{{\color{black} #1}}

\newcommand{\name}{\emph{SMooGPT}\xspace}

\definecolor{llightgray}{RGB}{230,230,230}

\acmJournal{TOG}

\begin{document}

\title{\name: Stylized Motion Generation using Large Language Models}

\author{Lei Zhong}
\affiliation{%
 \institution{University of Edinburgh}
 \streetaddress{10 Crichton Street}
 \city{Edinburgh}
 \country{United Kingdom}
 }
\email{zhongleilz@icloud.com}

\author{Yi Yang}
\affiliation{%
 \institution{University of Edinburgh}
 \streetaddress{10 Crichton Street}
 \city{Edinburgh}
 \country{United Kingdom}
 }
 \email{Y.Yang-249@sms.ed.ac.uk}

\author{Changjian Li}
\affiliation{%
 \institution{University of Edinburgh}
 \streetaddress{10 Crichton Street}
 \city{Edinburgh}
 \country{United Kingdom}
 }
 \email{chjili2011@gmail.com}

\begin{teaserfigure}
\centering
\begin{overpic}[width=0.98\linewidth]{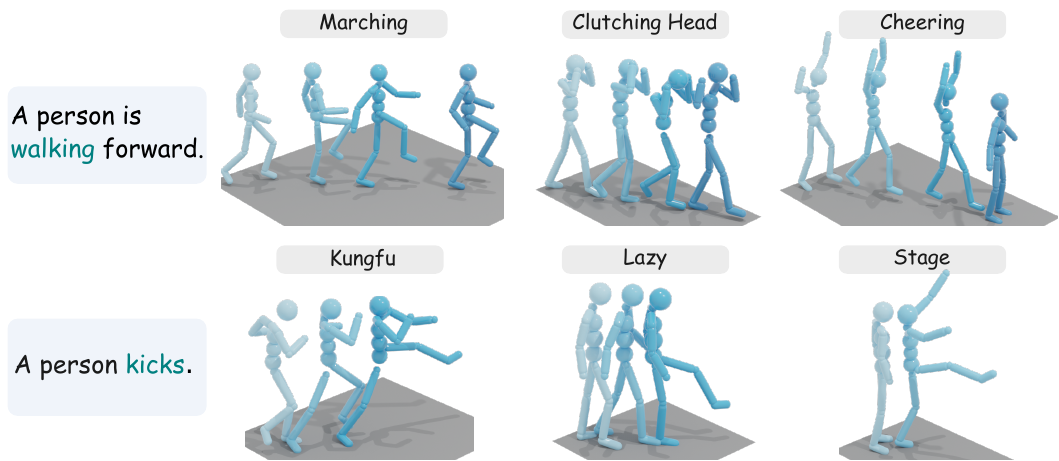}
\end{overpic}
\caption{\textbf{Stylized Motion Generation.} Given a motion \emph{content text} (left) and a variety of motion \emph{style texts} (top), we present a novel approach for generating stylized motions that concretely express the styles through the generated motion. Our method is highly versatile, supporting not only textural inputs but also motion sequences-or a combination of both-as motion content or style. See the paper, supplemental material, and video for more vivid examples.
}
\label{fig:teaser}
\end{teaserfigure}

\begin{abstract}
Stylized motion generation is actively studied in computer graphics, especially benefiting from the rapid advances in diffusion models. The goal of this task is to produce a novel motion respecting both the motion \emph{content} and the desired motion \emph{style}, \eg ``Walking in a loop like a Monkey''.
Existing research attempts to address this problem via motion style transfer or conditional motion generation. They typically embed the motion style into a latent space and guide the motion implicitly in a latent space as well. Despite the progress, their methods suffer from low interpretability and control, limited generalization to new styles, and fail to produce motions other than ``walking'' due to the strong bias in the public stylization dataset. 
In this paper, we propose to solve the stylized motion generation problem from a new perspective of reasoning-composition-generation, based on our observations: 
\rev{i) human motion can often be effectively described using natural language in a body-part centric manner,
ii) LLMs exhibit a strong ability to understand and reason about human motion,}
and iii) human motion has an inherently compositional nature, facilitating the new motion content or style generation via effective recomposing. 
We thus propose utilizing body-part text space as an intermediate representation, and present \name, a fine-tuned LLM, acting as a reasoner, composer, and generator when generating the desired stylized motion.
Our method executes in the body-part text space with much higher interpretability, enabling fine-grained motion control, effectively resolving potential conflicts between motion content and style, and generalizes well to new styles thanks to the open-vocabulary ability of LLMs.
Comprehensive experiments and evaluations, and a user perceptual study, demonstrate the effectiveness of our approach, especially under the pure text-driven stylized motion generation.

\end{abstract}



\maketitle

\section{Introduction}
Human motion can generally be decomposed into two components: content and style.
The content describes what the movement entails (\eg walking, waving), whereas the style characterizes how the movement is executed, often reflecting personality traits (\eg elderly, childlike) or emotional states (\eg happy, angry). 
To automatically and efficiently generate stylized motion, existing works~\cite{jang2022motion,wen2021autoregressive,aberman2020unpaired} have primarily focused on transferring style from an input stylized motion sequence to a target content motion sequence.
Recent advances in text-to-motion diffusion models~\cite{tevet2023human,chen2023executing} enable direct stylized motion generation~\cite{zhong2024smoodi,sawdayee2025dance,guo2025stylemotif,li2024mulsmo} using both motion sequences and texts as content and style inputs.%

Although the above works have made significant progress, a key limitation is that they typically embed the motion style into a latent space, and then either fuse it with content motion features (\eg \cite{wen2021autoregressive}) or inject it into a pretrained generative model (\eg SMooDi~\cite{zhong2024smoodi}) to produce stylized motion.
However, decoupling and recomposing style and content within the latent space might cause conflicts between content and style (\eg arms from ``Aeroplane'' and ``Throwing a ball''), and often lack interpretability.
Moreover, these methods are usually constrained by styles available in curated datasets (\eg 100STYLE~\cite{mason2022local} on all kinds of ``walk''-alike motions), leading to a strong bias toward the action ``walk'' 
and the limited generalization ability to new motion styles.

\begin{figure}[!t]
    \centering
    \begin{overpic}[width=0.97\linewidth]{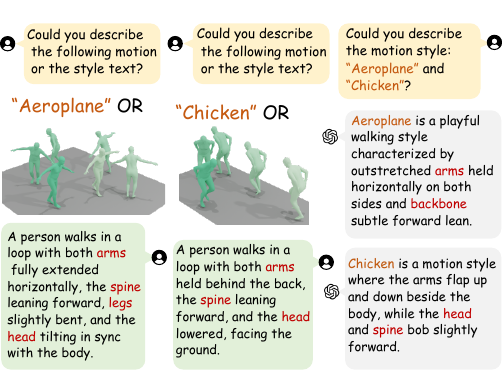} %
        \put(14, -3.5) {\small (a) }
        \put(47, -3.5) {\small (b)}
        \put(83, -3.5) {\small (c)}
        \put(18, 71) {Human Interpretation}
        \put(72, 71) {LLM Reasoning}
    \end{overpic}
    \caption{\textbf{Motion, Body Parts, and Natural Language.}
    Human motion is a coherent composition of body-part movements that can be described well using natural language. (a) A style motion can be naturally characterized by a combination of body parts styles, and similarly, (b) arbitrary content motion can be depicted by the same set of body parts using natural language. (c) LLMs have the intrinsic power to understand human motion and produce part-based descriptions in line with human interpretation in the language domain (see the highlighted common color).
    }
    \vspace{-4mm}
    \label{fig:insights_motion_text}
\end{figure}
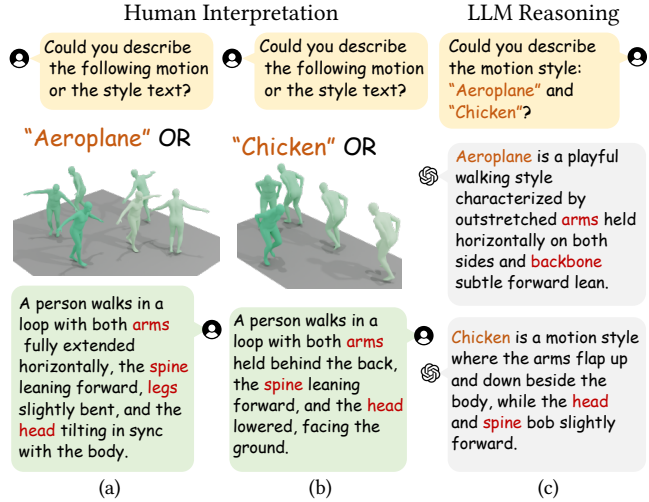

In this paper, we propose solving this motion stylization problem from a new perspective of body part motion analysis.
Our key insight is that both motion content and style can be interpreted as fine-grained control over body-part movements, and these 3D motions can be further intuitively expressed with natural language.
For example, as shown in Fig.~\ref{fig:insights_motion_text}(a) and (b), the ``Aeroplane'' style is characterized primarily by the upward extension of both arms, while the ``Chicken'' motion can be interpreted with the same set of body parts in the \emph{text} domain.
Given the intrinsic correlation among motion, body parts, and natural language, we investigate a fundamental research question: \emph{can stylized motion generation be effectively achieved through body-part centric textual descriptions}?

\rev{
At the same time, Large Language Models (LLMs) exhibit strong capabilities in understanding and reasoning about human poses~\cite{feng2024chatpose,li2025unipose} and motions~\cite{lin2024chathuman,li2025chatmotion,chen2025motionllm} (see an example in Fig.~\ref{fig:insights_motion_text}(c))—including stylized ones—and their representational scope extends well beyond that of existing motion style datasets~\cite{mason2022local,xia2015realtime}. Crucially, recent work~\cite{li2025much} empirically confirms that LLMs possess the capability to accurately deconstruct complex motion instructions into part-specific descriptions. 
Capitalizing on this unique advantage, it is compelling to leverage the learned motion knowledge and reasoning abilities of LLMs to decompose and recompose motion style and content within the \emph{text} domain. This enables more interpretable and conflict-free compositional representations of stylization.
}
On the other hand, interpreting stylized motion at the body-part level enables more effective use of existing datasets. While a new motion style may be unseen at the global level, many of its constituent body-part motions are likely to already exist within the dataset.

Building upon these insights, we propose \name, a fine-tuned LLM, to solve the stylized motion generation problem following a novel reasoning-composition-generation methodology. 
The core of our method is to harness the reasoning capabilities of LLMs to translate the desired motion content and style into body-part texts, which are then used to generate the desired motion.
Specifically, given the motion content and style represented as motion sequences or texts, \name is first instructed to reason out the corresponding body-part texts for each of them, acting as a motion reasoner.
Next, instead of directly combining body-part texts from the content and style, which might potentially cause conflict, \name is further instructed to compose the two sets of body-part texts into unified and conflict-free body-part texts that respect both the content and the style, acting as a composer.
Finally, given the composed body-part texts, \name is instructed to generate the expected stylized motion, acting as a motion generator.
To enable the versatile abilities, we have fine-tuned an LLM with pre- and post-training stages on curated datasets derived from HumanML3D~\cite{Guo_2022_CVPR} and 100STYLE~\cite{mason2022local}.
The goal of pre-training is to bridge the text and motion modalities, enabling motion understanding and reasoning \rev{in the text domain}, while the post-training is to enable composition and generation.
\rev{Furthermore, a lightweight diffusion head is leveraged to resolve inconsistencies between independent body-part motions, ensuring a harmonized and coordinated final stylized motion.
}
By leveraging the understanding and reasoning capabilities of LLMs within the \emph{body-part text domain}, our approach effectively resolves conflicts between motion content and style, and demonstrates a certain degree of generalization to new styles and contents - challenges that remain difficult for state-of-the-art methods at present.

We compare our approach on HumanML3D~\cite{Guo_2022_CVPR} and 100STYLE~\cite{mason2022local} benchmarks against SoTA methods, achieving superior performance, especially on the pure text-driven setting. Extensive qualitative experiments, ablation studies, and discussions validate our approach's effectiveness and scalability. A further user evaluation also demonstrates our advantages.
Additionally, by operating within the body part-based textual domain, our approach supports not only fine-grained control over body parts but also highly flexible and versatile stylized motion generation, where motion sequences and text descriptions can serve interchangeably as content or style inputs, enabling a wide range of applications. 

In summary, the principal contributions of this work include:
\begin{itemize}
    \item We propose an \emph{reasoning-composition-generation} methodology for stylized motion generation, utilizing the reasoning abilities of LLMs to understand and compose human motion in the text domain.

    \item We propose utilizing body-part text as an intermediate representation. By translating content and style requirements into this text domain, our approach effectively alleviates the challenge of content-style decoupling and composing, and facilitates the generation in an interpretable manner.

    \item Experiments demonstrate that \name not only sets a new state of the art in stylized motion generation, but also enables generalization to styles beyond existing motion datasets.
\end{itemize}

\section{Related Work}

We review recent advancements in human motion generation, motion stylization, and large language models (LLMs) and multi-modal LLMs (MLLMs) in the human motion area. 

\paragraph{Human Motion Generation}
Human motion synthesis focuses on generating diverse and realistic human-like movements.
Traditional approaches rely on concatenation and retrieval techniques, including motion graphs~\cite{kovar2023motion,heck2007parametric,shin2006fat} and motion matching~\cite{buttner2015motionmatching}.
Recently, learning-based methods have been widely adopted to generate high-quality and realistic motions, utilizing models such as GANs~\cite{li2022ganimator,ghosh2021synthesis}, autoregressive GPTs~\cite{guo2022tm2t,jiang2023motiongpt,zhang2023generating,zhangmotiongpt}, generative masked Transfomers~\cite{pinyoanuntapong2024mmm,guo2024momask,li2024lamp,meng2025rethinking}, and diffusion models~\cite{tevet2023human,motiondiffuse,zhou2025emdm,zhong2025sketch2anim}. %

\paragraph{Motion Stylization}
Motion stylization~\cite{aberman2020unpaired,kim2024most,tang2023rsmt,wen2021autoregressive,mason2022real,wu2025semantically,chen2025clusterstyle} is the process of creating stylized motion based on content and style specifications.
Early methods mainly focus on motion style transfer~\cite{song2024arbitrary,tao2022style,aberman2020unpaired}, typically leveraging existing content and style sequences for recombination.
Specifically, Aberman \etal~\shortcite{aberman2020unpaired} proposed the use of AdaIN to separate motion style from content and enable their flexible re-composition, which has since been widely adopted by many follow-up works~\cite{guo2024generative}.
Motion Puzzle~\cite{jang2022motion} realizes a framework that can control the style of individual body parts.
A major limitation of these models is their reliance on specialized style datasets~\cite{mason2022real,xia2015realtime} with limited
motion content, which restricts their applications

More recently, to extend the expressiveness of motion content, some approaches~\cite{zhong2024smoodi,guo2025stylemotif,li2024mulsmo,sawdayee2025dance} leverage pre-trained text-to-motion models~\cite{tevet2023human,chen2023executing} to generate stylized motions from text and style sequences.
Specifically, SMooDi~\cite{zhong2024smoodi} customizes a pre-trained text-to-motion model on the 100STYLE dataset to enable stylization across diverse motion content and styles.
StyleMotif~\cite{guo2025stylemotif} and MulSMo~\cite{li2024mulsmo} extend the style input to a multimodal setting based on SMooDi
\footnote{Code from StyleMotif and MulSMo was not publicly available at the time of submission.}.
However, these methods still face challenges in generalizing to diverse motion styles that are unseen in the dataset.
In contrast, we utilize the body-part textual domain and propose an \emph{reasoning-compositing-generation} framework, which harnesses the reasoning ability of LLMs, ensuring the generalization of our approach to unseen styles in the wild.

\paragraph{Multimodal Language Models}
Large Language Models~\cite{devlin2019bert,dai2019transformer,raffel2020exploring,brown2020language,zhang2022opt,touvron2023llama} have shown remarkable capabilities in textual comprehension and reasoning. 
For instance, 
T5~\cite{raffel2020exploring} presents a unified solution by converting all language-related tasks into a text-to-text formulation.
In a later stage, the pre-trained LLMs are adapted for multimodal tasks in various domains - 
mesh generation~\cite{wang2024llama}, garment generation~\cite{bian2024chatgarment}, and motion generation~\cite{shan2025mojito,wu2024motion,yu2025socialgen,li2024unimotion}.

More relevant to our scenario, 
MotionGPT~\cite{jiang2023motiongpt,zhangmotiongpt,zhu2025motiongpt3} treats human motion as a foreign language and leverages the language understanding and transfer abilities of pre-trained language models to support various tasks, like motion in-betweening, text-to-motion generation, and motion-to-text captioning. It also establishes the popular methodology of motion tokenization and LLM-based interaction.
Based on this methodology, MotionChain~\cite{jiang2024motionchain} supports multi-turn interactions by sampling single-turn data into multi-turn training examples, while Chen \etal~\shortcite{chen2024body_of_language} includes more modalities (\eg emotion, audio, and hands) in the interaction with human motion in LLMs.
Although our technical implementation is similar to MotionGPT~\cite{jiang2023motiongpt}, an obvious distinction is that we introduce the body-part textual domain and take it as a common intermediate representation to achieve fine-grained content and style control in the stylized motion generation task. 
Besides, none of the above models can naturally handle this task, and \name complements existing work.

\begin{figure}[!t]
    \centering
    \begin{overpic}[width=0.90\linewidth]{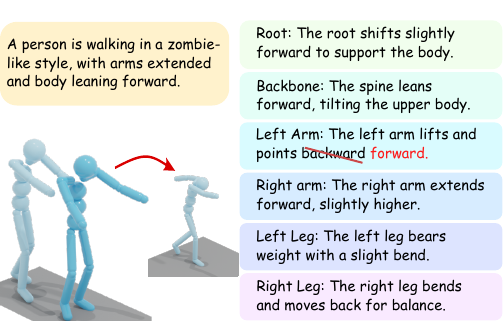} %
        \put(2,61.5) {\small Global Text $\mathbf{T}_g$}
        \put(49,61.5) {\small Body-part Text $\{\mathbf{T}_p\}$}
        \put(2, 38) {\small 4D motion $\mathbf{m}$}
        \put(24,34) {\footnotesize Editing}
    \end{overpic}
    \caption{\textbf{Body Part Space.} 
    Our body-part space consists of three main elements - the global text, the body-part texts, and the corresponding 4D motion. By changing from ``backward'' to ``forward'' of the left arm, a novel stylized motion can be obtained (\ie ``Superhero''), due to the compositional nature of the space (bottom-left).
    }
    \vspace{-4mm}
    \label{fig:data_cons}
\end{figure}

\begin{figure*}[!t]
    \centering
        \begin{overpic}[width=0.97\textwidth]{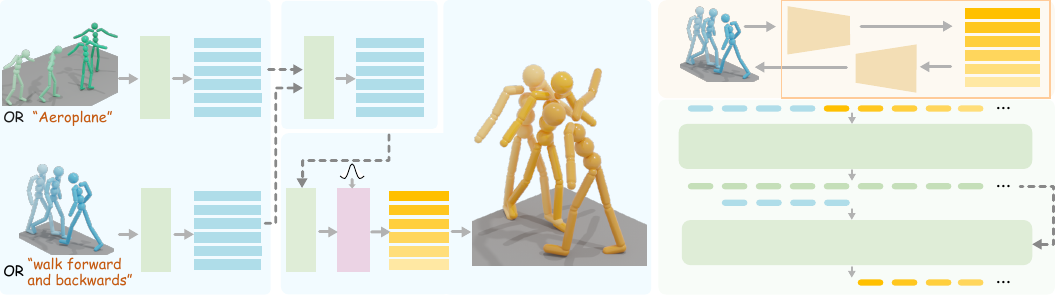} %
        \put(8.5,26.5) {\small (i) Reasoning}
        \put(29.2,26.5) {\small (ii) Composing}
        \put(47.2,26.5) {\small (iii) Generation}

        \put(32.9, 15.6) {\footnotesize Body-part text}
        \put(17.4, 15.6) {\footnotesize Body-part text}
        \put(17.4, 0.7) {\footnotesize Body-part text}
        \put(35.8, 0.7) {\footnotesize Motion token}

        \put(34.8, 11.3) {\footnotesize Noise}

        \put(14.0,17){\rotatebox{90}{\name}}
        \put(29.5,17){\rotatebox{90}{\name}}
        \put(14.1,2.5){\rotatebox{90}{\name}}
        \put(27.9,2.5){\rotatebox{90}{\name}}
        \put(32.3,3.5){\rotatebox{90}{\footnotesize Diffusion}} 
        \put(33.5,4.5){\rotatebox{90}{\footnotesize Head}}
        \put(77,13.5){{\large Encoders}}
        \put(71.3,4.5){{\large Autoregressive Decoders}}
        \put(75.2,23.8){\footnotesize \shortstack{ Motion\\Encoder}}
        \put(81.8,20.2){\footnotesize\shortstack{Motion\\Decoder}}
        \put(84.5,26.5){\small $\times6$}
        \put(64.8,8.3){\footnotesize <bos>}

        \put(63,18.5) {\small (Sec. \ref{subsec:tokenization})}
        \put(63,1) {\small (Secs. \ref{subsec:pre_training}\&\ref{subsec:post_training})}
        
        \put(22,-1.8) {\small (a) Motion Generation Overview}
        \put(75,-1.8) {\small (b) LLM Finetuning}
    \end{overpic}
    \caption{\textbf{Stylized Motion Generation.}
   (a) Our body-part centric stylized motion generation methodology adopts a reasoning, composing, and generation framework, where the given motion content and style (either texts or motion sequences) are translated and composed in the body-part space, and the resulting body-part texts are further translated back to the motion space, producing the stylized motion. (b) We invent \name by fine-tuning an LLM with body part-based tokenization (top), pre- and post-training (top) stages, operating in our methodology. 
    }
    \vspace{-2mm}
    \label{fig:pipeline}
\end{figure*}
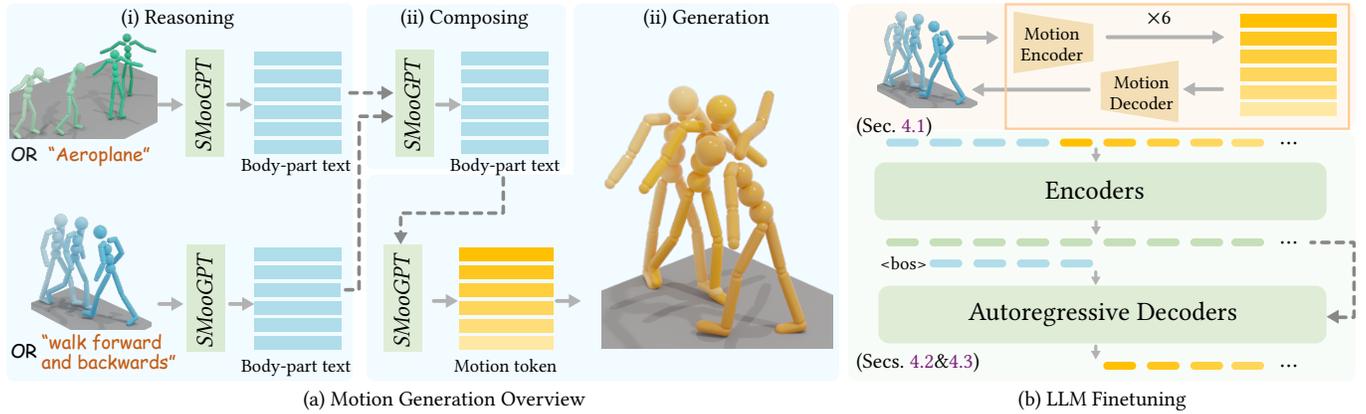

\section{Motion Generation Overview}
\label{sec:methodology}

\subsection{Body-part Space}
Stylized motion is inherently compositional, with stylistic attributes often manifested unevenly across different body parts. While global motion representations offer a high-level summary, they fail to explicitly capture localized variations.
Introducing a body-part space enables a more structured and localized representation of motion, allowing fine-grained control of individual body parts. This formulation enhances the interpretability of stylized motion by disentangling how different body parts contribute to the overall style.

Formally, our body-part space is formulated as a triplet $\{\mathbf{m}, \mathbf{T}_g, \{\mathbf{T}_p\}\}$.
As illustrated in Fig.~\ref{fig:data_cons}, each motion $\mathbf{m}$ in the body-part space can be interpreted as a composition of part-specific behaviors collectively expressing the full-body movement. 
The global text $\mathbf{T}_g$ primarily captures the overall semantics of the motion along with high-level stylistic modifiers (e.g., ``walk forward and then walk backward in a flying pose''), while the body-part texts $\{\mathbf{T}_p\}$ provide detailed, localized descriptions for individual body parts.
Importantly, motion style is manifested not only in static poses but also in dynamic movement patterns. Therefore, our body-part textual descriptions are designed to capture both the spatial configuration of each body part and its dynamic properties, such as movement direction and rhythm. 
To demonstrate the compositional nature of our body-part representation, Fig.~\ref{fig:data_cons} (bottom-left) presents an example in which modifying only a few body-part descriptions—\eg changing the left arm motion from pointing  ``backward'' to ``forward''—produces a new stylized motion while preserving the underlying content. This example underscores the flexibility of our design: part-level representations improve the reusability of existing datasets and enable compositional generation, allowing novel motion styles to be synthesized by recombining previously observed body-part movements.

\subsection{Generation Methodology}
\label{subsec:methodology}

Fig.~\ref{fig:pipeline}(a) demonstrates an overview of our generation pipeline, composed of three main components - motion reasoning, composition, and generation, powered by our fine-tuned LLM - \name.

Specifically, users interact with \name in the textual domain and execute the three steps sequentially. In the following, we demonstrate the instructional interactions:

\begin{itemize}
    \item Reasoning: given the input motion content or style in the format of motion sequence or text, we instruct \name to output the body-part texts. \textbf{Note that} if the input is a motion sequence $\mathbf{m}$, an extra embedding step is applied (see Sec.~\ref{subsec:tokenization}).

\begin{lstlisting}[label={lst:reasoning_prompt}]
Input: (*@$\mathbf{m}$@*) or <(*@$\mathbf{T}_g$@*)>
Task_Prompt: Reason out the body-part texts
Answer: Root: <(*@$\mathbf{T}_{root}$@*)>, Backbone: <(*@$\mathbf{T}_{bb}$@*)>, Left Arm: <(*@$\mathbf{T}_{la}$@*)>, Right Arm: <(*@$\mathbf{T}_{ra}$@*)>, Left Leg: <(*@$\mathbf{T}_{ll}$@*)>, Right Leg: <(*@$\mathbf{T}_{rl}$@*)>.
\end{lstlisting}

    \item Composition: given the two body-part text sets, we explicitly ask \name to compose them to obtain unified and conflict-free body-part texts, representing the desired body configuration and motion.
\begin{lstlisting}[label={lst:composition_prompt}]
Input: Content body-part texts <(*@$\{\mathbf{T}_{p}\}$@*)>, and Style body-part texts <(*@$\{\mathbf{T}_{p}\}$@*)>
Task_Prompt: Compose the two body-part text sets into a unified and conflict-free body-part text set.
Answer: Root: <(*@$\mathbf{T}_{root}$@*)>, Backbone: <(*@$\mathbf{T}_{bb}$@*)>, Left Arm: <(*@$\mathbf{T}_{la}$@*)>, Right Arm: <(*@$\mathbf{T}_{ra}$@*)>, Left Leg: <(*@$\mathbf{T}_{ll}$@*)>, Right Leg: <(*@$\mathbf{T}_{rl}$@*)>.
\end{lstlisting}

    \item Generation: having the unified body-part texts, the last task is to simply translate them into the motion space, producing the resulting motion. \rev{Note that the body-part specific motion tokens will first go through a diffusion denoising process (Sec. \ref{subsec:diffusion_denoising}) to obtain the harmonized and coordinated motion tokens, which are further decoded back to the motion space using the pre-trained decoders (see Sec.~\ref{subsec:tokenization}).}
\begin{lstlisting}[label={lst:generation_prompt}]
Input: Body-part texts <(*@$\{\mathbf{T}_{p}\}$@*)>
Task_Prompt: Given the body-part texts, generate the corresponding body-part motions
Answer: Root: <motion_tokens>, Backbone: <motion_tokens>, Left Arm: <motion_tokens>, Right Arm: <motion_tokens>, Left Leg: <motion_tokens>, Right Leg: <motion_tokens>.
\end{lstlisting}

\end{itemize}

\section{Method}
\label{sec:method}

To gain the abilities acting as the above reasoner, composer, and generator, we fine-tune an LLM on a curated dataset, as shown in Fig.~\ref{fig:pipeline}(b).
We elaborate on details in the following.

\subsection{Motion Tokenization}
\label{subsec:tokenization}

To better align with the LLM's next-token prediction mechanism and to incorporate body-part level reasoning, we partition the whole-body motion into multiple part motions.
Specifically, given a motion sequence \( \mathbf{m} \in \mathbb{R}^{N \times H} \), where \( N \) is the number of frames, and \( H \) is the feature dimension. We employ the same motion representations as in HumanML3D~\cite{Guo_2022_CVPR}, where $H=263$.
We decompose \( \mathbf{m} \) into six body-part motions \( m_p \in \mathbb{R}^{N \times H_p} \), where \( p \in \{\text{Right Arm}, \text{Left Arm}, \text{Right Leg}, \text{Left Leg}, \text{Backbone}, \text{Root}\} \) \cite{zou2024parco}.
For each part motion $m_{p}$, a part-specific encoder $\mathcal{E}_{p}$ maps it into a latent representation $\hat{c}_p=\mathcal{E}_{p}({m}_p)$, which is then quantized by finding the nearest code in a part-specific codebook $\mathcal{C}_p$ consisting of K learnable latent vectors in $\mathbb{R}^d$. The quantization process is formulated as:
\begin{equation}
c_p = \mathrm{Quantize}(\hat{c}_p) = \arg\min_{c_p^k \in \mathcal{C}_p} \| \hat{c}_p - c^k_p \|_2.
\end{equation}
The paired part-specific decoder $\mathcal{D}_p$ subsequently reconstructs ${m}_p=\mathcal{D}_p(c_p)$.

The VQ-VAE for each body part is trained to minimize the following objective:
\begin{equation}
\mathcal{L}_{\text{VQ}}^p = \| {m}_p - \mathcal{D}_p(c_p) \|_2^2 + \| \mathrm{sg}[\mathcal{E}_p({m}_p)] - c_p \|_2^2 + \beta \| \mathcal{E}_p({m}_p) - \mathrm{sg}[c_p] \|_2^2,
\end{equation}
where \( \mathrm{sg}[\cdot] \) represents the stop-gradient operator, and \( \beta \) is the weight for the commitment loss.
This part-wise tokenization provides synchronized discrete token sequences across different body parts, facilitating fine-grained motion modeling and reasoning in the subsequent stages.

\subsection{Pre-training for Modality Alignment}
\label{subsec:pre_training}

In this stage, body-part motions are regarded as a foreign language and are integrated into the token space of the pre-trained LLM. We then learn motion and language simultaneously in a unified vocabulary. 
Concretely, we expand the original text-specific vocabulary $\mathcal{B}_t$ (\ie WordPiece tokens) by introducing body-part motion tokens $\mathcal{B}_p = \{c_p^{1:K}\}$, where $p$ belongs to the above pre-defined body parts, and $c_p^i$ is the motion token in the corresponding codebook.
\rev{For example, each motion token is prefixed with the body-part name (e.g., \texttt{left\_arm\_<idx>}).}
Moreover, we insert special tokens $\mathcal{B}_{s}$ (\eg </som\_lr> and </eom\_lr> as the start and end indicators of the left arm motion) into the LLM's vocabulary to indicate the beginning and end of each body-part motion sequence.
Finally, the original vocabulary is extended into a unified text-motion vocabulary, $\mathcal{B}=\{\mathcal{B}_{t},\mathcal{B}_{p},\mathcal{B}_{s} \}$.

Equipped with the vocabulary $\mathcal{B}$, we align body-part motion and language through bidirectional translation tasks.
Specifically, for each data sample in our dataset, we have the body-part space triplet representation, \ie $\{\mathbf{m}, \mathbf{T}_g, \{\mathbf{T}_p\}\}$. With the help of the tokenization step above, the triplet can be written as:
\begin{equation*}
    \left\{ \mathbf{m}= \{m_p\} \approx \{c_p\}, \mathbf{T}_g, \{\mathbf{T}_p\} \right\},
\end{equation*}
and the translation happens between $\{c_p\}$ and $\{\mathbf{T}_p\}$. The task follows the format of ``predicting body-part texts from body-part motions'' or ``predicting body-part motions from body-part texts''.

\subsection{Post-training with Instruction Following}
\label{subsec:post_training}

After pre-training, the model captures the underlying grammatical structure of the body-part motion vocabulary and establishes strong alignment between motion and text modalities.

As shown in Fig.~\ref{fig:pipeline}(a) and stated in Sec.~\ref{subsec:methodology}, our key idea is to fully exploit the intermediate body-part text space for stylized motion generation. 
To this end, we design several instruction-following templates for the above bidirectional translations, as well as the extra global text to body-part texts predictions

With the aforementioned pre- and post-training on our style-agnostic dataset, \name becomes a general-purpose motion reasoner and generator that operates on body-part texts.
As for the composition ability, we design an additional instruction following task, where we fine-tune \name with content body-part texts $\mathbf{T}_p^c$ and style body-part texts $\mathbf{T}_p^s$ as shown below. Note that superscripts $c$ and $s$ differentiate the content text from the style text, and please refer to the supplementary to know their construction.

\subsection{Language Model Training Details}
\label{subsec:training_detaiils}

Our model leverages 220M pre-trained Flan-T5-Base model~\cite{raffel2020exploring} with an encoder–decoder transformer structure to address the conditional generation task.
Through our extended vocabulary, both body-part textual descriptions and motion tokens are represented uniformly as "text" tokens.
Specifically, both the input and output can be converted into a sequence of \( S_{i/o} = \{ s_{i/o}^k \}_{k=1}^{L} \), where \( s_{i/o} \in \mathcal{B} \) and \( L \) represent the sequence length.

Since our model is an encoder-decoder architecture, we set a maximum input length of 512. Following the original T5 implementation, the sequence of tokens is sent to the encoder, and the decoder then performs next-token prediction in an autoregressive manner at each step. The training objective in both pre-training and post-training can be formulated as follows:
\begin{equation}
\mathcal{L}_{LM} = - \sum_{k=0}^{L_t - 1} \log p_\theta(s_o^k \mid s_o^{<k}, s_i).
\end{equation}
For both pre- and post-training, we finetune the model's entire weights. 
More training details can be found in the \emph{supplementary}.

\rev{
\subsection{Diffusion-based Motion Refinement}
\label{subsec:diffusion_denoising}
To bridge the gap between discrete LLM outputs and continuous motion dynamics, we introduce a lightweight diffusion head preceding the decoder.
Specifically, the LLM-predicted body-part latents $\{c_p\}$ are aggregated into a unified condition vector $\bm{z}_{\text{cond}}$. 
We then employ a conditional denoiser $\epsilon_\theta(\bm{z}_t, t, \bm{z}_{\text{cond}})$ to iteratively refine a latent vector starting from Gaussian noise $\bm{z}_T \sim \mathcal{N}(\bm{0}, \bm{I})$. 
This process yields a globally coordinated latent $\bm{z}_0$, which is subsequently decomposed back into body-part features and decoded by the VQ-decoder $\mathcal{D}$ to generate the final smooth motion sequence $\bm{m}$.
}

\begin{figure*}[!t]
    \centering
    \includegraphics[width=0.95\textwidth]{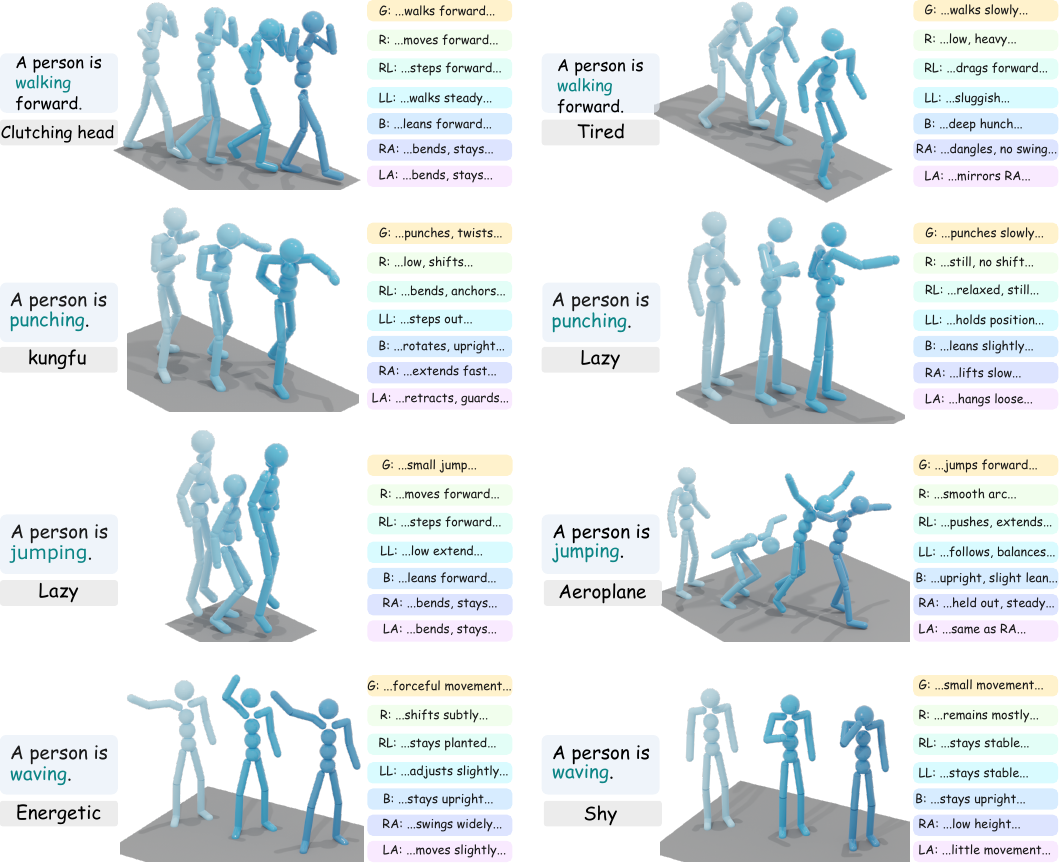}
    \caption{\textbf{Result Gallery.}
    Using \name, we have generated a diverse set of stylized motions guided by various combinations of motion content and style texts. To highlight our method's capabilities, we focus on specialized pure text-based inputs. Each example presents the content and style descriptions, the resulting motion depicted through selected frames, and the corresponding simplified body-part texts. The full body-part texts are available in the supplementary. 
    }
    \label{fig:gallery}
\end{figure*}

\begin{figure*}[!t]
    \centering
    \includegraphics[width=\textwidth]{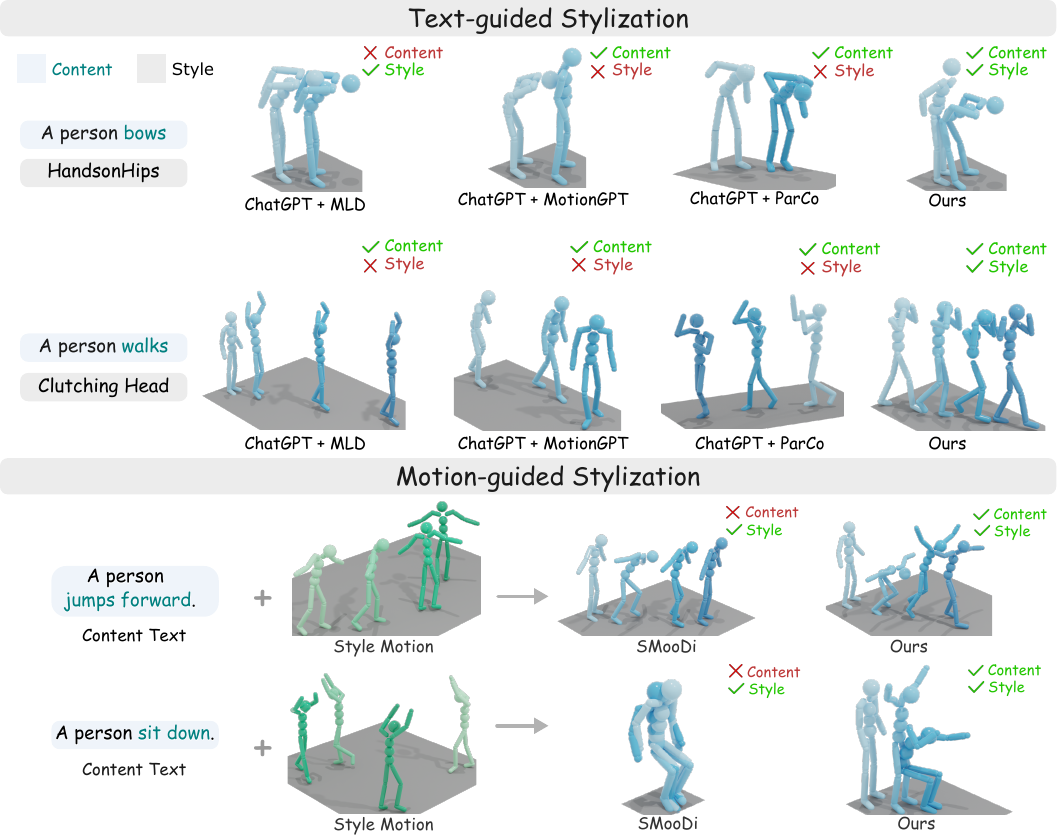}
    \caption{\textbf{Visual Comparison.} Typical examples of the comparison between our method and competitors on the text-guided and motion-guided stylization tasks are shown. Pay attention to the imperfection in either the motion content or the style of the generated motion from competitors.
    }
    \label{fig:comparison}
\end{figure*}

\begin{figure*}[!t]
    \centering
    \includegraphics[width=1.0\textwidth]{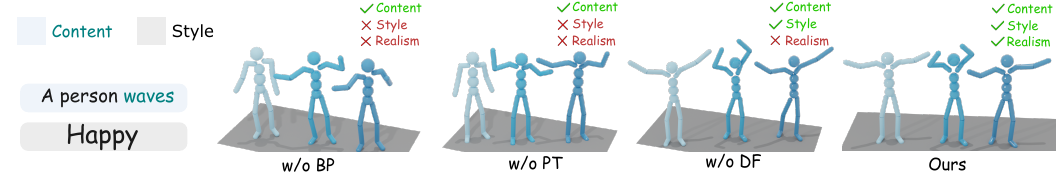}
    \caption{\textbf{Visual Ablation.} Qualitative comparison of ablated variants under the same content and style texts. Removing the body-part space (w/o BP), pre-training (w/o PT), or diffusion head (w/o DF)
leads to degraded style expression, realism, or content--style consistency,
while the full model produces coherent and realistic stylized motion.
Please refer to the accompanying video for detailed comparisons.
    }
    \label{fig:ablation}
\end{figure*}

\section{Results and Evaluation}
\label{sec:results}

With our \name, we have generated many appealing stylized motions, ranging from ``Kungfu kick'', ``Aeroplane jumping'', to ``Lazy punch'', as presented in Figs.~\ref{fig:teaser} and \ref{fig:gallery}. 
Note that, styles like ``Marching'', ``Stage Performance'', ``Cheering'', are not labeled or explicitly present in our HumanML3D training data, which reflects our method's ability to generalize to unseen or underrepresented styles, based solely on their natural language descriptions, benefiting from LLM’s text reasoning ability.
We also show simplified body-part texts in Fig.~\ref{fig:gallery}, and the full version can be found in the supplementary.
Although our method is versatile in stylized motion generation by accepting both motion sequences and text descriptions as conditions, we specialize in pure \emph{text-guided motion stylization}. 
Thus, we mainly demonstrate this application in the gallery (Fig.~\ref{fig:gallery}).
Additionally, we compare baseline methods from motion-guided stylization in Sec.~\ref{subsec:comp}.
Ablation studies are conducted in Sec.~\ref{subsec:abl_study}, validating our technical design. 
A user perceptual study (Sec.~\ref{subsec:user_eval}) also supports our superior performance.
\rev{Some critical discussions are presented in Sec.~\ref{subsec:discussion} and the \emph{supplementary}.}
A supplemental video is provided for better visualization of the motion. 

\paragraph{Dataset.}
We choose HumanML3D dataset~\cite{Guo_2022_CVPR} as our \emph{training} set for both the pre- and post-training. We utilize its consistent root-velocity motion representation and process the dataset to obtain our body-part space representation.
For \emph{evaluation}, we use the HumanML3D test set as the content source and randomly pair styles represented as texts or motion sequences from the test set of 100STYLE dataset~\cite{mason2022local}. 
The detailed data construction process is in the supplementary.

\begin{table*}[!t]
\caption{\textbf{Quantitative Comparison Results}. We compare with baselines on text-guided stylization.
Motion content preservation (MM Dist $\downarrow$, R-Precision (Top-3) $\uparrow$, CLIP-scoreea $\uparrow$), realism (FS-Ratio $\downarrow$), and style reflection (SRA-L $\uparrow$) metrics are reported. The best results are highlighted in \textbf{bold}.}
\label{tab:compare_main}

\vspace{-3mm}
\centering
\resizebox{0.80\textwidth}{!}{
\begin{tabular}{lccccc}
\toprule
\textbf{Method} 
& \textbf{SRA-L} $\uparrow$ 
& \textbf{MM Dist} $\downarrow$ 
& \textbf{R-Precision (Top-3)} $\uparrow$ 
& \textbf{CLIP-score} $\uparrow$
& \textbf{FS-Ratio} $\downarrow$ \\

\midrule
ChatGPT + MotionGPT~\cite{jiang2023motiongpt} 
& 11.562  & 1.225 & 0.373 & 0.614 & 0.0984 \\
ChatGPT + ParCo~\cite{zou2024parco} 
& 13.359  & 1.212 & 0.405 & 0.622 & 0.111 \\

ChatGPT + MLD~\cite{chen2023executing} 
& 9.609  & 1.246 & 0.289 & 0.652 & 0.181 \\
\textbf{SMooGPT (Ours)} 
& \textbf{17.344} & \textbf{1.132} & \textbf{0.669} & \textbf{0.670} & \textbf{0.0832} \\

\bottomrule
\vspace{-3mm}
\end{tabular}
}
\end{table*}

\paragraph{Evaluation metrics.}
We evaluate our method along three dimensions: \emph{content preservation}, \emph{style reflection}, and \emph{realism}.
For the assessment of content preservation, we employ metrics used in~\cite{chen2023executing}: motion-retrieval precision (R-Precision), Multi-modal Distance (MM Dist). 
Moreover, we utilize the CLIP-score~\cite{meng2025rethinking} to evaluate the alignment between the text content and the stylized motion.
Additionally, we exploit the foot skating ratio (FS-Ratio) proposed by~\cite{karunratanakul2023gmd} for our motion realism evaluation.
For style reflection, we employ Style Recognition Accuracy (SRA)~\cite{jang2022motion}.
\rev{Since the style classifier is trained on the 100STYLE dataset—which primarily consists of locomotion—it may be unreliable for non-locomotion content. Consequently, we calculate SRA by filtering for locomotion-related sequences within the HumanML3D test set, a metric we denote as SRA-L.}
We do not include the common FID metric for evaluating content preservation due to its inherent ambiguity in our task. \emph{See supplementary for the detailed discussion.}

\subsection{Comparison}
\label{subsec:comp}

\paragraph{Text-guided Stylization.}
This task generates stylized motion based on both the content and style texts.
\rev{All methods were trained only on the HumanML3D dataset, where styles are not labeled or explicitly present.
Two concurrent works, StyleMotif~\cite{guo2025stylemotif} and MulSMo~\cite{li2024mulsmo}, are capable of producing stylized motion in this setting. However, both methods rely on style motion sequences from the 100STYLE dataset to align style labels. Furthermore, their source code was not publicly available at the time of submission.}
We thus create three baseline methods by integrating a pre-trained LLM (\ie GPT-3.5~\cite{openai2024chatgptgeneral}) with three representative motion generators (\ie MotionGPT~\cite{jiang2023motiongpt}, ParCo~\cite{zou2024parco}, and MLD~\cite{chen2023executing}).
Strategically, we utilize content text from the HumanML3D dataset and style labels from the 100STYLE dataset.
We first ask ChatGPT to combine the content and style label, and employ the respective motion generator to produce the corresponding motion according to the composed text.
Statistical results are displayed in Tab.~\ref{tab:compare_main}. 

In terms of style reflection (SRA-L), \name outperforms the strongest baseline by approximately \textbf{29.8\%}. Furthermore, our method achieves superior performance across all motion content preservation and realism metrics. Specifically, we improve content consistency by reducing MM-Dist by \textbf{6.6\%} and increasing R-Precision by \textbf{65.2\%}, while also achieving the best motion realism with the lowest FS-Ratio (\textbf{0.0832}) among all compared methods.
The corresponding visual results are presented in Fig.~\ref{fig:comparison}(top). In three typical cases, baseline methods either cannot preserve the content or fail to reflect the style.

\begin{table*}[!t]
\caption{\textbf{Quantitative Comparison} on motion-guided stylization. 
Motion content preservation (MM Dist $\downarrow$, R-Precision (Top-3) $\uparrow$, CLIP-scoreea $\uparrow$), realism (FS-Ratio $\downarrow$), and style reflection (SRA-L $\uparrow$) metrics are reported.
The best results are highlighted in \textbf{bold}.}
\label{tab:compare_motion_guided}
\centering
\vspace{-2mm}
\renewcommand{\arraystretch}{1.2} 

\resizebox{0.65\textwidth}{!}{
\begin{tabular}{lccccc}
\toprule
\textbf{Method} 
& \textbf{SRA-L} $\uparrow$ 
& \textbf{MM Dist} $\downarrow$ 
& \textbf{R-Precision (Top-3)} $\uparrow$ 
& \textbf{CLIP-score} $\uparrow$
& \textbf{FS-Ratio} $\downarrow$ \\

\midrule
\rowcolor{gray!15} \multicolumn{6}{c}{\textbf{Setting 1: Zero-Shot (Cross-Dataset)}} \\

SMooDi~\cite{zhong2024smoodi} 
& 7.110 & 1.185 & \textbf{0.504} & 0.640 & 0.112 \\
\textbf{SMooGPT (Ours)} 
& \textbf{12.695} & \textbf{1.153} & 0.481 & \textbf{0.659} & \textbf{0.104} \\ 

\midrule
\rowcolor{gray!15} \multicolumn{6}{c}{\textbf{Setting 2: Held-Out (Intra-Dataset)}} \\ 

SMooDi~\cite{zhong2024smoodi} 
& 14.775 & 1.157 & \textbf{0.509} & 0.651 & 0.118 \\
\textbf{SMooGPT (Ours)} 
& \textbf{15.527} & \textbf{1.150} & 0.474 & \textbf{0.660} & \textbf{0.110} \\ 

\bottomrule
\end{tabular}
}
\vspace{-3mm}
\end{table*}

\paragraph{Motion-guided Stylization}
This task generates stylized motion from a content text and a reference style motion.
To the best of our knowledge, SMooDi~\cite{zhong2024smoodi} is the only method designed for this setting, and thus serves as our primary baseline.
To evaluate generalization to unseen styles, we consider two evaluation protocols.
The \textbf{zero-shot (cross-dataset)} setting trains both \name and SMooDi on HumanML3D and evaluates them using content texts from HumanML3D paired with style motions from 100STYLE.
The \textbf{held-out (intra-dataset)} setting reserves $10$ styles from 100STYLE for testing, ensuring that target styles are unseen during training.
To ensure a fair comparison under unseen settings, we adapt the configurations of both models.
Specifically, \name directly incorporates the reference style motion into the diffusion process to avoid information loss from text translation (see supplementary),
while SMooDi disables its classifier-based style guidance to avoid reliance on style supervision requiring seen styles.

The statistical and visual comparisons are presented in Tab.~\ref{tab:compare_motion_guided} and Fig.~\ref{fig:comparison} (bottom), respectively.
In the \textbf{zero-shot (cross-dataset)} setting, \name achieves better performance than SMooDi in all metrics other than R-Precision.
Specifically, our method achieves a remarkable \textbf{78.5\%} improvement in style reflection. 
This validates that our strategy of performing composition in the body-part text space confers superior generalization capabilities for stylization.
In the \textbf{held-out (intra-dataset)} setting, our approach maintains this advantage, surpassing SMooDi in both style expression and motion realism.
Despite slightly lower R-Precision in two settings, \name consistently outperforms SMooDi on other content preservation metrics, including MM Dist and CLIP-score.

\begin{table}[!t]
\caption{\textbf{Statistical Ablation Studies Results}.}
\label{tab:abl_studies}

\centering
\vspace{-3mm}
\renewcommand{\arraystretch}{1.2}
\resizebox{1.0\linewidth}{!}{
\begin{tabular}{lccccc}
\toprule
\textbf{Method} 
& \textbf{SRA-L} $\uparrow$ 
& \textbf{MM Dist} $\downarrow$ 
& \textbf{R-P.(Top-3)} $\uparrow$ 
& \textbf{CLIP-score} $\uparrow$
& \textbf{FS-Ratio} $\downarrow$ \\
\midrule

w/o BP 
& 13.674 & 1.231 & 0.452 & 0.6324 & 0.0963 \\
w/o PT
& 15.234 & 1.380 & 0.501 & 0.667 & 0.0958 \\
w/o DF 
& 16.328 & 1.134 & 0.512 & 0.669 & 0.0953 \\
\rowcolor{llightgray}
\textbf{Ours} 
& \textbf{17.344} & \textbf{1.132} & \textbf{0.520} & \textbf{0.670} & \textbf{0.0832} \\

\bottomrule
\end{tabular}
}
\vspace{-4mm}
\end{table}

\subsection{Ablation Study}
\label{subsec:abl_study}
To validate the effectiveness of each module in our framework, we conduct several ablation studies on the text-guided stylization task and show the visual ablation in Fig.~\ref{fig:ablation}.

\paragraph{The importance of body-part space.}
To assess the impact of incorporating the body-part text space, we conduct an ablation study using the original MotionGPT~\cite{jiang2023motiongpt} model—referred to as w/o BP—which is trained solely on global textual inputs within the same dataset. 
As illustrated in Tab~\ref{tab:abl_studies}, incorporating body-part information yields a \textbf{26.8\%} improvement in style reflection, demonstrating the advantage of fine-grained control over individual body parts.
Moreover, this structured representation simultaneously enhances semantic alignment—boosting R-Precision by \textbf{15.0\%} and CLIP-score by \textbf{5.9\%}—while significantly improving motion realism by reducing the foot sliding ratio by \textbf{13.6\%}.

\paragraph{The importance of pre-training.}
To assess the role of pre-training, we remove the pre-training phase (denoted as w/o PT).
As shown in Tab.~\ref{tab:abl_studies}, this leads to consistent performance degradation across all metrics, including a \textbf{12.2\%} drop in style reflection and a substantial \textbf{21.9\%} increase in MM Dist.
These results highlight the importance of pre-training for aligning fine-grained body-part texts with motion representations, which in turn facilitates effective post-training and improves overall performance.

\paragraph{The importance of the diffusion head.}
To evaluate the contribution of the diffusion head, we perform an ablation study by excluding this component (denoted as w/o DF). As shown in Tab.~\ref{tab:abl_studies}, incorporating the diffusion head yields substantial performance gains in both realism and stylization.
Specifically, the inclusion of the diffusion head reduces the FS-Ratio by approximately \textbf{12.7\%}.
These improvements validate that the diffusion head plays a critical role in coordinating body-part motions, ultimately leading to more harmonious motion.

\subsection{User Evaluation}
\label{subsec:user_eval}
\begin{figure}[!t]
    \centering
    \begin{overpic}[width=0.99\linewidth]{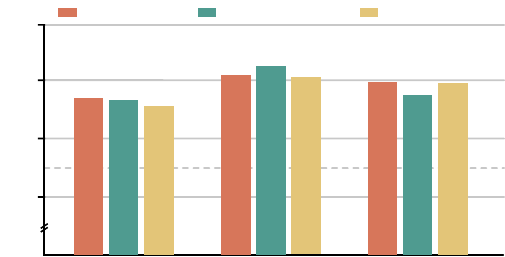} %
        \put(-2, 9) {\rotatebox{90}{\footnotesize \name wins (\%)}}

        \put(3.5, 10.8) {\footnotesize 40}
        \put(3.5, 22.2) {\footnotesize 60}
        \put(3.5, 33.5) {\footnotesize 80}
        \put(2, 45) {\footnotesize 100}
        
        \put(76, 47.5)  {\scriptsize Content Preservation}
        \put(44, 47.5)  {\scriptsize Style Reflection}
        \put(16, 47.5)  {\scriptsize Motion Realism}
        
        \put(14.4, 32)  {\footnotesize 74.5}
        \put(21.6, 31.8)  {\footnotesize 73.5}
        \put(28.8, 31.0)  {\footnotesize 72.5}
        
        \put(44.3, 36.5)  {\footnotesize 81.3}
        \put(51.0, 38)  {\footnotesize 81.7}
        \put(58.3, 35.9)  {\footnotesize 80.7}
        
        \put(73, 35)  {\footnotesize 79.57}
        \put(80.3, 32.4)  {\footnotesize 75.1}
        \put(87, 35)  {\footnotesize 79.53}

        \put(23, -4)  {\small (a)}
        \put(52, -4)  {\small (b)}
        \put(81, -4)  {\small (c)}

    \end{overpic}
    \caption{\textbf{User Evaluation Statistics.}
    The percentage of times our approach is preferred over (a) ChatGPT+MLD, (b) ChatGPT+MotionGPT, and (c) ChatGPT+ParCo is reported. When choosing from a pair of generated motions, users are asked to evaluate three aspects - Content Preservation, Style Reflection, and Motion Realism. The higher the percentage (50\% is a tie), the better our results.
    }
    \label{fig:user_study}
\end{figure}

We conduct a user perceptual study using pairwise comparisons on the text-guided stylization task.
In each trial, participants are shown a motion content text, a motion style text, and two motion clips generated by \name and a competing method (ChatGPT+MLD, ChatGPT+MotionGPT, or ChatGPT+ParCo), and are asked to evaluate content preservation, style reflection, and motion realism (Fig.~\ref{fig:user_study}).
We invited $75$ participants in total, and Fig.~\ref{fig:user_study} reports the percentage of times \name is preferred over competing methods.

Overall, \name is consistently favored across all three aspects.
Specifically, it achieves preference rates of $81.3\%$, $81.7\%$, and $80.7\%$ against ChatGPT+MotionGPT, remains above $72\%$ when compared with ChatGPT+MLD, and obtains similarly strong results against ChatGPT+ParCo ($79.57\%$, $75.1\%$, and $79.53\%$).
Overall, the user perceptual study validates our superior performance and is consistent with observations in the comparison.

\begin{figure}[!t]
    \centering
    \begin{overpic}[width=0.99\linewidth]{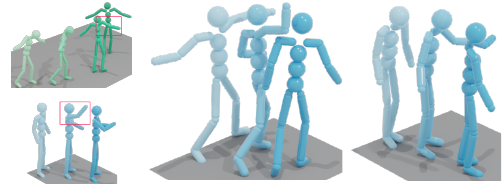}
        \put(4.5, 18.7)  {\small Style Motion}
        \put(4.5, -0.5)  {\small Content Motion}
        \put(46.5, -0.5)  {\small Ours}
        \put(81, -0.5)  {\small SMooDi}
    \end{overpic}
    \caption{\textbf{Conflict Resolving.}
    (a) Given the motion content ``A person with right hand throwing'' and the motion style  ``Aeroplane'', (b) our method produces plausible stylized motion, resolving the body-part conflicts (see the highlighted boxes) and respecting both aspects, while (c) SMooDi, taking the style motion as input, fails to address it.
    }
    \label{fig:conflict_resolving}
\end{figure}

\subsection{Discussions}
\label{subsec:discussion}

\paragraph{The importance of composition.}
Fueled by the reasoning capabilities of LLMs, our approach acts as an effective composer that reconciles motion content and style texts in a conflict-free manner, which is critical when the two impose contradictory constraints on specific body parts.
Fig.~\ref{fig:conflict_resolving} shows a representative example: the motion content ``throw'' requires one arm to stay close to the body, while the ``Aeroplane'' style demands both arms be lifted and extended outward.
Our method resolves this conflict by selectively borrowing one arm configuration from each source, producing a composed motion where one arm performs the throwing action and the other adopts the aeroplane pose.
In contrast, SMooDi, which takes style motion rather than style text as input, fails to coherently satisfy both content and style.

\paragraph{Why do we not use FID for evaluating content preservation?}
FID measures the overall distributional similarity between generated and real motion samples in a latent feature space. While it is useful for assessing global motion realism, it is agnostic to the input conditioning. 
In our setting, stylized motions are generated based on explicit content and style inputs. 
The style component intentionally alters the motion distribution, which causes the FID to increase, even when the content is faithfully preserved. 
As a result, a high FID is ambiguous: it is difficult to determine whether it reflects successful style modulation over preserved content or a failure to express both content and style correctly. 
Therefore, FID cannot reliably reflect whether the generated motion matches the intended content.

\section{Conclusion}
We propose a new paradigm for stylized motion generation by framing it as a reasoning–composition–generation process in a body-part-centric text space.
By leveraging the semantic richness and reasoning ability of large language models (LLMs), our \name, a fine-tuned LLM, achieves interpretable, controllable, and generalizable stylized motion synthesis that surpasses prior methods in handling novel style. 
Extensive experiments and user studies validate the effectiveness of our method. We believe our work opens a new direction for motion generation by unifying natural language understanding with structured motion representation.

\bibliographystyle{ACM-Reference-Format}
\bibliography{main}

\clearpage
\section*{Supplemental Material}

\appendix

\setcounter{table}{0}
\renewcommand{\thetable}{A\arabic{table}}
\setcounter{figure}{0}
\renewcommand{\thefigure}{A\arabic{figure}}

In this supplementary material, we provide additional details that support and complement the main paper.
Specifically, we include detailed descriptions of dataset construction, implementation details of our method and all baselines, and extended examples of body-part text representations corresponding to the generated motions.
We provide a supplemental video, which we encourage the reviewers to watch since motion is critical in our results, and this is hard to convey in a static document.

\section{More Discussions}
\label{subsec:more_discussion}

\paragraph{Semantic composition vs. latent disentanglement.}
Unlike prior methods that rely on explicit feature disentanglement constraints to separate style from content—a process often fraught with optimization instability—we propose a paradigm shift towards implicit semantic composition. By projecting both motion content and style into a unified, interpretable body-part text space, we bypass the need for complex losses or specialized architectures to enforce disentanglement. Our framework leverages the reasoning capabilities of LLMs to merge these attributes at the linguistic level, ensuring that style and content are naturally integrated into conflict-free body-part instructions before generation.s

\paragraph{LLM-driven motion priors and body-part decomposition}
The core assumption of \name is that LLMs possess inherent knowledge of human motion and motion styles, alongside the capability to decompose complex movements into body-part-level components. Based on this, we establish a reason-compose-generation pipeline that leverages LLMs for semantic reasoning.
Recent research~\cite{li2025much} provides strong empirical evidence for this approach. Their findings confirm that LLMs excel at high-level movement planning and can accurately deconstruct complex instructions into part-specific descriptions. 
This directly validates our "part-based decomposition" strategy.
Furthermore, they demonstrate that LLMs exhibit robust "motion commonsense" for creative and culturally specific styles that are often underrepresented in traditional mocap datasets. 
This finding provides critical support for \name's assumption regarding LLMs' inherent knowledge of motion styles.

\paragraph{Style and Body-Part Alignment Capability.}
To assess whether large language models can reason about the relationship between motion styles and the body parts they primarily affect, we conduct a quantitative evaluation on the 100STYLE dataset.
Since the original dataset does not provide explicit body-part annotations for styles, we manually annotate each style by inspecting representative motion sequences and identifying the body parts that exhibit the most prominent stylistic variations.
Each style is annotated with one or more body-part labels, forming a multi-label ground-truth set.

For evaluation, we prompt GPT using only the global style name and its textual description, without providing any motion data, and ask it to predict the relevant body parts from a predefined list.
We then compare the predicted body-part sets with the human-annotated ground truth using the micro-F1 score under a multi-label classification setting.
The resulting micro-F1 score of $0.72$ indicates a strong alignment between the LLM predictions and human annotations, supporting our claim that LLMs possess a meaningful prior for reasoning about style--body-part relationships in structured motion control.

\paragraph{The coordination of body-part motion.}
\name utilizes two coordination mechanisms: (1) implicit coordination, occurring during the LLM-driven translation from body-part text to motion tokens, and (2) explicit coordination, in which the diffusion head directly models the dependencies between body-part motion tokens.

\section{Instruction Templates for Post-training}
\label{sec:supp_inst}

This section provides the instruction templates used for post-training with instruction following, as described in Sec.~4.3 of the main paper.

\paragraph{Global-to-Body-Part Decomposition.}
This instruction trains the model to decompose a global motion description into fine-grained body-part-centric texts, where each text describes the localized motion behavior of a specific body part.

\begin{lstlisting}[label={lst:tuning_prompt1}]
Input: Global motion text <(*@$\mathbf{T}_g$@*)>
Task_Prompt: Given <(*@$\mathbf{T}_g$@*)>, generate a detailed body-part centric text description, explaining the described motion. The body parts are ..., following the order.
Answer: Root: <(*@$\mathbf{T}_{root}$@*)>, Backbone: <(*@$\mathbf{T}_{bb}$@*)>, Left Arm: <(*@$\mathbf{T}_{la}$@*)>, Right Arm: <(*@$\mathbf{T}_{ra}$@*)>, Left Leg: <(*@$\mathbf{T}_{ll}$@*)>, Right Leg: <(*@$\mathbf{T}_{rl}$@*)>.
\end{lstlisting}

\paragraph{Content--Style Composition.}
This instruction trains the model to compose motion content and motion style descriptions in the body-part space, resolving potential conflicts between them to produce a unified and coherent description.

\begin{lstlisting}[label={lst:tuning_prompt2}]
Input: Motion content body-part texts <(*@$\{\mathbf{T}_p^c\}$@*)> and motion style body-part texts <(*@$\{\mathbf{T}_p^s\}$@*)>
Task_Prompt: Given the content and style body-part texts, generate a unified and coherent body-part text set, which should explain both the motion and style harmoniously. Pay attention to the potential conflicts between content and style.
Answer: Root: <(*@$\mathbf{T}_{root}$@*)>, Backbone: <(*@$\mathbf{T}_{bb}$@*)>, Left Arm: <(*@$\mathbf{T}_{la}$@*)>, Right Arm: <(*@$\mathbf{T}_{ra}$@*)>, Left Leg: <(*@$\mathbf{T}_{ll}$@*)>, Right Leg: <(*@$\mathbf{T}_{rl}$@*)>.
\end{lstlisting}

\section{Dataset Constructions}

To operationalize the overall pipeline, we construct a body-part-level motion–text dataset that supports reasoning, composition, and generation.
We collect two types of data: 
\begin{enumerate}
    \item reasoning from content and style texts to body-part-level textual annotations; 
    \item composing the content and style body-part texts into a unified body-part description that captures only the pose and dynamic movement.
\end{enumerate} 

To support (1), we primarily rely on the HumanML3D~\cite{Guo_2022_CVPR} and 100STYLE~\cite{mason2022local} datasets, which provide annotated global content texts and style labels, respectively. 
\rev{The HumanML3D dataset contains 14,646 motions with 44,970 text annotations. In contrast, 100STYLE~\cite{mason2022local}, the largest motion style dataset, offers 1,125 minutes of motion sequences covering 100 diverse locomotion styles.}
We employ ChatGPT-3.5~\cite{openai2024chatgptgeneral} to decompose these global descriptions into body-part-specific textual annotations. 
\rev{Specifically, we use HumanML3D to generate content texts for individual body parts and leverage 100STYLE to obtain part-level style descriptions that characterize how each body part moves according to its style labels.}
For (2), we further utilize the 100STYLE dataset to construct unified body-part texts that jointly encode both content and style. Following the SMooDi~\cite{zhong2024smoodi} protocol, we first annotate each motion in 100STYLE with a global content text by MotionGPT~\cite{jiang2023motiongpt}, then prompt ChatGPT-3.5 to infer body-part-level content descriptions. 
Finally, we prompt ChatGPT-3.5 to reason about the relationship between body-part content and style texts. It first synthesizes a global description that briefly captures the overall motion, followed by unified textual representations for each individual body part.
This unified textual representation captures both static poses and dynamic movements of each body part—without relying on any explicit global style description.

\paragraph{Held-out style subset construction.}
To enable a fair comparison between SMooGPT and SMooDi under the held-out setting, we partition the 100STYLE dataset into seen and held-out style subsets. 
Since some style labels in 100STYLE inherently encode content-specific cues, we primarily follow the style grouping protocol adopted in SMooDi to avoid content leakage.
Specifically, we designate the following styles as held-out and exclude them entirely from training: 
\texttt{Zombie}, \texttt{SwingShoulders}, \texttt{SwingArmsRound}, \texttt{Swimming}, \texttt{Swat}, \texttt{Superman}, \texttt{Stiff}, \texttt{Star}, \texttt{ShieldedRight}, and \texttt{ShieldedLeft}.
During stylization-enhanced tuning, the model is trained only on the remaining seen styles and is evaluated on the held-out styles to assess generalization.

\rev{For out-of-domain evaluation, we randomly sample text or motion sequences from the out-of-domain subsets as style inputs and use the model trained on the in-domain subsets to assess generalization.}

\section{Implementation Details}
\paragraph{Evaluation Metrics.} Given that the standard evaluator~\cite{Guo_2022_CVPR} can bias embeddings toward text alignment at the expense of motion fidelity, we instead utilize TMR~\cite{petrovich2023tmr} to compute MM-Dist and R-Precision. This choice is consistent with recent protocols~\cite{hwang2025snapmogen} to ensure a more robust assessment of content preservation.

\paragraph{Body-part VQ-VAE}
Our approach employs six lightweight VQ-VAEs to discretize part-specific motions for each body part.
The architecture and training methodology largely follow ParCo~\cite{zou2024parco}.
Each VQ-VAE employs a codebook with 512 entries. All body parts are encoded using 128-dimensional codes, except for the \textit{Root} part, which uses a 64-dimensional code. The downsampling rate is set to \( r = 4 \).
Further implementation details can be found in ParCo~\cite{zou2024parco}.

\rev{
\paragraph{Diffusion-based Motion Refinement}
While LLMs can generate conflict-free body-part texts and translate them into motion tokens, this process operates sequentially in a discrete space and lacks explicit modeling of continuous coordination across body parts. To bridge this gap, we introduce a lightweight diffusion head preceding the decoder.
Let $\mathcal{C}_{in} = \{c_p\}_{p=1}^P$ denote the set of independent body-part motion latents predicted by the LLM. These inputs are first projected and composed into a unified continuous representation $\bm{z}_{\text{cond}} = \Psi(\mathcal{C}_{\text{in}})$ using a composition function $\Psi$ implemented as a set of MLPs followed by a concatenation operation.

The core of this module is the denoiser $\epsilon_\theta$ parameterized by $\theta$. Architecturally, we implement $\epsilon_\theta$ as a set of MLPs followed by a concatenation operation.
The diffusion process operates on a sequence of unified motion latents $\{\bm{z}_t\}_{t=0}^{T}$, starting from Gaussian noise $\bm{z}_T \sim \mathcal{N}(\bm{0}, \bm{I})$.
At each denoising step $t$, the model predicts the noise component to refine the latent:
\begin{equation}
    \epsilon_t = \epsilon_\theta(\bm{z}_t, t, \bm{z}_{\text{cond}}).
\end{equation}
We optimize the model using the standard noise reconstruction loss:
\begin{equation}
    \mathcal{L}_\text{recon} = \mathbb{E}_{\epsilon, \bm{z}_0, t} \left[ \left\| \epsilon_{\theta}(\bm{z}_t, t, \bm{z}_{\text{cond}}) - \epsilon \right\|^2_2 \right],
\end{equation}
where $\epsilon \sim \mathcal{N}(\mathbf{0}, \mathbf{I})$ represents the ground-truth noise added to the clean latent $\bm{z}_0$.

After $T$ iterative denoising steps, a clean and coordinated unified latent $\bm{z}_0$ is obtained. Crucially, applying the inverse operation $\Psi^{-1}$ decomposes $\bm{z}_0$ back into body-part representations, resulting in $\{\hat{c}_p\}_{p=1}^{P}$. Finally, these refined body-part latents are decoded by $\mathcal{D}$ into the final motion sequence $\bm{m} \in \mathbb{R}^{N \times H}$.
}

\paragraph{Motion-guided stylization.}
In this task, we aim to synthesize stylized motion conditioned on both content text and a style motion sequence.
A straightforward approach involves using \name to first translate the style sequence into body-part textual descriptions, compose them with the content text, and subsequently generate the output.
However, projecting high-frequency style motion into the discrete text domain inevitably results in a loss of fine-grained dynamic details.
To mitigate this, drawing inspiration from SMooDi~\cite{zhong2024smoodi}, we employ a lightweight adapter to inject style guidance into the diffusion head.
Specifically, the denoising step is modified to condition on the style embedding. Let $\bm{s}$ denote the style features extracted from the reference motion $\bm{m}_r$. The noise prediction network $\epsilon_\theta$ is updated to:

\begin{equation}
    {\epsilon}_t = \epsilon_\theta(\bm{z}_t, t, \bm{z}_{\text{cond}}, \bm{s}).
\end{equation}
The standard noise reconstruction loss is updated to:
\begin{equation}
    \mathcal{L}_\text{recon} = \mathbb{E}_{\epsilon, \bm{z}_0, t} \left[ \left\| \epsilon_{\theta'}(\bm{z}_t, t, \bm{z}_{\text{cond}},\bm{s}) - \epsilon \right\|^2_2 \right].
\end{equation}
In the zero-shot setting, models are trained solely on HumanML3D. Conversely, in the intra-dataset setting, training is performed on HumanML3D augmented with the seen subset of 100STYLE.

\paragraph{Pretraining and instruction following}
We leverage T5~\cite{chung2024scaling} as the underlying architecture for our language model, with 12 layers in both the Transformer encoder and decoder. 
The model is trained with a learning rate of \(2 \times 10^{-4}\) during the pre-training stage and \(1 \times 10^{-4}\) during the instruction tuning stage, using a mini-batch size of 48 for both. 
Each stage consists of 300K training iterations.
To prevent the model from over-relying on the global text or specific body parts, we randomly mask out either the global text or one to two body-part texts during training.
For the stylization-enhanced tuning, the model is trained for an additional 50K iterations. 
All models are trained on 2 NVIDIA H200 GPUs.

\paragraph{Inference time}
To evaluate the inference efficiency of our approach, we report the average inference time of generating a single motion clip, measured in seconds, on a single NVIDIA RTX4090 GPU.
\rev{Specifically, our approach takes approximately 25.761 seconds to generate a motion clip.}

\section{Design Details of Competitors}
\paragraph{Text-guided motion stylization}
Since no existing framework directly addresses the problem of text-guided motion stylization, we develop baseline approaches that integrate ChatGPT with existing motion generation methods, MotionGPT~\cite{jiang2023motiongpt}, MLD~\cite{chen2023executing} and ParCo~\cite{zou2024parco}.
Specifically, we follow the same procedure used in dataset construction to compose the content text and style label into a unified global description, which is then fed into the three motion generation models.

\paragraph{Motion-guided motion stylization}
We leverage the source code of SMooDi to retrian in the two settings.
For the zero-shot setting, we early-stop training at $10$ epochs to prevent content forgetting caused by over-focusing on the reference motion.
For the held-out setting, we follow the original SMooDi training protocol and train the model for $50$ epochs.
To balance the trade-off between content preservation and style reflection, we set the classifier-free guidance weight to $1$. Furthermore, we disable classifier-based guidance during inference, as it relies on a pre-trained encoder that fails to generalize to unseen reference motions.

\begin{figure}[!t]
    \centering
    \includegraphics[width=\linewidth]{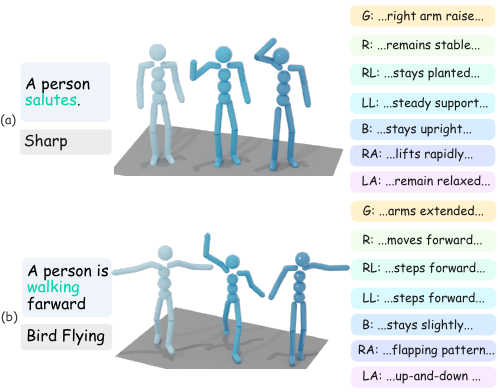}
    \caption{\textbf{Failure Cases.}
    Representative failure cases illustrating the limitations of our method.
    \emph{Left}: For motions requiring sharp temporal transitions (e.g., a sharp salute), the model captures the overall arm trajectory but fails to produce an abrupt stopping phase.
    \emph{Right}: For out-of-distribution motion patterns (e.g., a bird-flying walk), the generated motion remains constrained to the learned motion manifold, resulting in missing periodic wing-flapping dynamics.}
    \vspace{-4mm}
    \label{fig:fail_cases}
\end{figure}

\section{User Evaluation Details}
We conduct a user perceptual study using pairwise comparisons to evaluate text-guided motion stylization.
In each test case, participants are presented with two motion clips generated by our method and a competing approach, respectively, given the same motion content text and motion style text.
To ensure a fair comparison, all videos are anonymized and randomized, and no information about the generating method is revealed to the participants.

Participants are asked to evaluate each motion pair based on three criteria: content preservation, style reflection, and motion realism.
For content preservation, participants assess how well the generated motion matches the semantics described in the motion content text.
For style reflection, they evaluate how clearly the target motion style is expressed in the generated animation.
For motion realism, participants judge the naturalness and physical plausibility of the motion, focusing on movement continuity and visual coherence.

A total of 21 motion pairs are evaluated, evenly divided into three groups corresponding to comparisons with three competing methods.
We invite 75 participants to complete the study, which is conducted online using Google Forms.
Each motion pair is presented side by side, and participants are required to indicate their preference for each of the three criteria independently.
\section{Limitations and Future Work}
While our method demonstrates strong performance, it remains subject to limitations imposed by the intermediate representation, as illustrated by the representative failure cases in Fig.~\ref{fig:fail_cases}.
First, the discrete nature of the body-part text space lacks the temporal granularity required to fully capture sharp and transient motion dynamics.
As shown in Fig.~\ref{fig:fail_cases}(a), although the model correctly generates the overall arm trajectory for a sharp salute, it fails to produce an abrupt stopping phase, resulting in an overly smooth transition near the motion peak.
This limitation arises from the fact that discrete, text-based descriptions cannot precisely specify fine-grained temporal phases such as instantaneous stops or rapid impulse changes.
Second, while the LLM exhibits strong reasoning capabilities for generalizing to novel styles, the final generation quality is fundamentally bounded by the representational capacity of the body-part VQ-VAE.
As illustrated in Fig.~\ref{fig:fail_cases}(b), for out-of-distribution motion patterns such as a bird-flying walk, the model captures the intended spatial configuration (e.g., extended arms) but fails to synthesize clear and periodic wing-flapping dynamics.
This behavior reflects the constraint that the decoder can only generate motions within the learned motion manifold, limiting its ability to extrapolate to motion patterns that are underrepresented or absent in the training data.

In future work, we aim to address these constraints by enriching the intermediate representation. Specifically, we plan to augment body-part textual descriptions with a series of keypose-based tokens that explicitly encode spatial configurations over time. Such temporally-aware extensions will enable the model to capture motion dynamics more effectively, offering a richer and more precise representation for both content and style.

\section{Additional Results}

\paragraph{Body-part space}
We showcase the detailed body-part text for the first row in the Gallery results in the following.

\begin{tcolorbox}[title=Case A, colback=gray!5!white, colframe=gray!75!gray, fonttitle=\bfseries]
\textbf{Global:} A person walks forward with both hands placed on their head.

\textbf{Root:} The root moves forward in a straight line with a small vertical oscillation, showing slightly reduced lift.

\textbf{Right Leg:} The right leg steps forward with moderate range, supporting the body’s forward movement.

\textbf{Left Leg:} The left leg follows with a regular pace, matching the rhythm of the walk.

\textbf{Backbone:} The spine leans forward slightly during the walk, with minimal upper body rotation.

\textbf{Right Arm:} The right arm bends at the elbow and remains positioned on the right side of the head with no swing.

\textbf{Left Arm:} The left arm bends at the elbow and remains positioned on the right side of the head with no swing.
\end{tcolorbox}

\begin{tcolorbox}[title=Case B, colback=gray!5!white, colframe=gray!75!gray, fonttitle=\bfseries]
\textbf{Global:} A person walks forward very slowly with their back deeply hunched, arms hanging heavily, and feet dragging across the ground.

\textbf{Root:} The root moves forward very slowly, staying low as if weighed down.

\textbf{Right Leg:} The right leg lifts barely off the ground and drags forward with each step.

\textbf{Left Leg:} The left leg follows with a sluggish motion, lacking any upward drive.

\textbf{Backbone:} The back is deeply hunched, with the upper body bending forward noticeably at the waist.

\textbf{Right Arm:} The right arm dangles low beside the body with no active swing.

\textbf{Left Arm:} The left arm mirrors the right, swaying slightly only due to momentum.
\end{tcolorbox}

\paragraph{Example: Global-to-Body-Part Decomposition.}
We provide a concrete example illustrating how the model decomposes a global motion description into fine-grained body-part-centric texts following the instruction template above.

\begin{flushleft}\small\ttfamily
Input (Global Motion Text):\\
"A person bends their left arm at the elbow and raises their right arm towards it, then lowers both arms."\\[0.5em]

Output (Body-Part Texts):\\
Root: The body remains mostly upright with a slight weight shift to maintain balance.\\
Backbone: The backbone stays straight and aligned, providing support for the arm movements.\\
Left Arm: The left arm bends at the elbow and serves as the focal point of the motion.\\
Right Arm: The right arm raises upward toward the left arm with controlled movement, then lowers.\\
Left Leg: The left leg provides stable support throughout the motion.\\
Right Leg: The right leg maintains a steady position with minimal movement.
\end{flushleft}

\paragraph{Example: Content--Style Composition.}
We provide an example illustrating how the model composes motion content and motion style descriptions in the body-part space to produce a unified and coherent representation.

\begin{flushleft}\small\ttfamily
Input (Motion Content Body-Part Texts):\\
Root: Moves forward with a steady walking motion.\\
Backbone: Remains upright to support forward locomotion.\\
Left Arm: Moves forward naturally as part of walking.\\
Right Arm: Moves forward naturally as part of walking.\\
Left Leg: Steps forward to initiate locomotion.\\
Right Leg: Steps forward to continue the walking motion.\\[0.5em]

Input (Motion Style Body-Part Texts: InTheDark):\\
Root: Moves cautiously with a restrained and steady gait.\\
Backbone: Maintains a controlled and composed posture.\\
Left Arm: Extends forward at chest height with minimal swing.\\
Right Arm: Extends forward at chest height with minimal swing.\\
Left Leg: Takes deliberate, measured steps.\\
Right Leg: Takes deliberate, measured steps.\\[0.5em]

Output (Composed Body-Part Texts):\\
Root: Moves forward with a steady and cautious gait, maintaining balance and coordination.\\
Backbone: Maintains stability and alignment throughout the motion.\\
Left Arm: Extends straight ahead at chest height, held steady while walking.\\
Right Arm: Extends straight ahead at chest height, held steady while walking.\\
Left Leg: Steps forward with a large diagonal movement to the front right.\\
Right Leg: Steps forward with a large diagonal movement to the front right.
\end{flushleft}

\end{document}